\documentclass[journal=ancac3,manuscript=article]{achemso}

\usepackage{chemformula} 
\usepackage[T1]{fontenc} 
\usepackage{siunitx}
\usepackage{hyperref}
\usepackage{amsmath}

\author{Stephen A. Church}
\email{stephen.church@manchester.ac.uk}
\author{Hoyeon Choi} 
\author{Nawal Al-Amairi}
\author{Ruqaiya Al-Abri}
\affiliation[the University of Manchester]
{Department of Physics and Astronomy and Photon Science Institute, the University of Manchester, Manchester M13 9PL, United Kingdom}
\author{Ella Sanders}
\affiliation[Weizmann Institute of Science]
{Department of Materials and Interfaces, Weizmann Institute of Science, Herzl St 234, Rehovot 7610001, Israel}
\author{Eitan Oksenberg}
\affiliation[AMOLF]
{Center for Nanophotonics, AMOLF, Amsterdam 1009 DB, the Netherlands}
\author{Ernesto Joselevich}
\affiliation[Weizmann Institute of Science]
{Department of Materials and Interfaces, Weizmann Institute of Science, Herzl St 234, Rehovot 7610001, Israel}
\author{Patrick W. Parkinson}
\email{patrick.parkinson@manchester.ac.uk}
\affiliation[the University of Manchester]
{Department of Physics and Astronomy and Photon Science Institute, the University of Manchester, Manchester M13 9PL, United Kingdom}

\title[Holistic Determination of Optoelectronic Properties using High-Throughput Spectroscopy of Surface-Guided CsPbBr3 Nanowires]
  {Holistic Determination of Optoelectronic Properties using High-Throughput Spectroscopy of Surface-Guided CsPbBr3 Nanowires}

\keywords{high-throughput, metal-halide perovskites, energy dynamics, photoluminescence, nanowires}

\begin{document}

\begin{abstract}
Optoelectronic micro- and nanostructures have a vast parameter space to explore for modification and optimisation of their functional performance. This paper reports on a data-led approach using high-throughput single nanostructure spectroscopy to probe $>$~8,000 structures, allowing for holistic analysis of multiple material and optoelectronic parameters with statistical confidence. The methodology is applied to surface-guided CsPbBr\textsubscript{3} nanowires, which have complex and interrelated geometric, structural and electronic properties. Photoluminescence-based measurements, studying both the surface and embedded interfaces, exploits the natural inter-nanowire geometric variation to show that increasing the nanowire width reduces the optical bandgap, increases the recombination rate in the nanowire bulk and reduces the rate at the surface interface. A model of carrier recombination and diffusion ascribes these trends to carrier density and strain effects at the interfaces and self-consistently retrieves values for carrier mobility, trap densities, bandgap, diffusion length and internal quantum efficiency. The model predicts parameter trends, such as the variation of internal quantum efficiency with width, which is confirmed by experimental verification. As this approach requires minimal \latin{a-priori} information, it is widely applicable to nano- and micro-scale materials.

\end{abstract}


Functional optoelectronic materials form the basis of a multitude of devices that are crucial to modern technology, including CCD detectors, photovoltaics, laser diodes and LEDs. There is particular interest in optoelectronic micro- and nano-structures, for increased density integration and enhanced performance with respect to planar devices. Some examples of these include micro-LED arrays for visible-light communication~\cite{Zhang2013} and nano-lasers~\cite{Duan2001,Kim2017} for on-chip photonic integrated circuits. In these structures, the functional performance is determined by many important optoelectronic properties~\cite{Alanis2019OpticalLasing}, such as the diffusion length, bandgap and defect density~\cite{Sergent2019}. The structural geometry, such as length, width or shape, is also an important factor, this can be highly coupled to other parameters, making nanotechnology targets particularly challenging to develop.

Automation and high-throughput spectroscopy offer a means to harness the variation in properties across a large population of nano- or microstructures. This fundamental approach has been used for materials spanning tens of microns in length, down to micron-scale\cite{Alanis2019OpticalLasing}. This allows correlations to be drawn between measured properties and can establish those with the greatest impact on performance~\cite{Parkinson2020}. In this study, this approach is applied to holistically characterise microstructures by studying each individual element with multiple techniques, including: photoluminescence spectroscopy, time-correlated single photon counting (TCSPC) and excitation power dependent TCSPC, with a typical characterisation time of a few seconds per NW. A self-consistent analysis is applied to this multi-modal data-set, of 15576 individual measurements, to correlate all of the measured properties and establish the coupling between geometry, strain and carrier recombination processes. This allows the extraction of important parameters with statistical confidence, such as the bandgap, diffusion length, trap densities and internal quantum efficiency (IQE). 

The multi-modal data-led approach is applied to surface-guided CsPbBr\textsubscript{3} nanowires (NWs). This material has recently been demonstrated to have bright luminescence~\cite{Oksenberg2021}, strong waveguiding properties~\cite{Shoaib2017} and low lasing thresholds~\cite{Schlaus2019}. These properties make the NWs ideal for applications in photodetectors~\cite{Zhou2018}, photovoltaics~\cite{Yuan2018} and on-chip coherent light sources~\cite{Eaton2016,Wang2018}. Optical techniques have previously been applied to study spatially resolved degradation and recombination and strain effects in thin films of halide perovskites~\cite{Choi2020,Frohna2021}. However, the optical behaviour of NWs demonstrate additional dependencies which increases the complexity of an experimental study. For example, the functional performance of these structures are strongly influenced by the strain~\cite{Sanders2021} and carrier trap densities, particularly at the surface, therefore fabrication processes are designed to control the material quality. However, these techniques do not control the NW geometry, and often lead to variation in dimensions on the single NW level~\cite{Oksenberg2018Surface-GuidedResponse}. The optoelectronic behaviour of these NWs is therefore described by a complex, correlated and multi-dimensional parameter space. The multi-modal data-led approach is required to gain an understanding of the properties of this challenging material system by harnessing this inter-NW variation.

\section{Results and Discussion}
\subsection{Experimental results}

CsPbBr\textsubscript{3} NWs were grown using a previously described recipe~\cite{Oksenberg2018Surface-GuidedResponse}. Scanning electron microscopy (SEM) was performed on a small number of the NWs (252): a subset of these are shown in Figure~1a. The mean NW length, and standard deviation (SD), was \SI{18(10)}{\micro\meter} and the mean width was $w = $~\SI{1.2(1)}{\micro\meter}. NWs grown with this approach have an approximately cubic crystal structure: the substrate/NW interface is the (110) plane, and the NW/air interface consists of both the (010) and (100) planes~\cite{Oksenberg2018Surface-GuidedResponse}. As a result, the NWs have an isosceles triangle cross section, i.e. with a height equal to half of the width. As discussed in ref \cite{Oksenberg2020}, the cubic lattice is slightly distorted due to strain and lattice rotation effects. The NW width was hence obtained independently using atomic force microscopy (AFM) on 116 NWs, where the mean height, and SD, was measured to be $h = w/2 =  $~\SI{0.6(2)}{\micro\meter}. More details are given in Figure S1 in the supporting information (S.I.).

\begin{figure*}
    \centering
    \includegraphics[width = 0.95\linewidth]{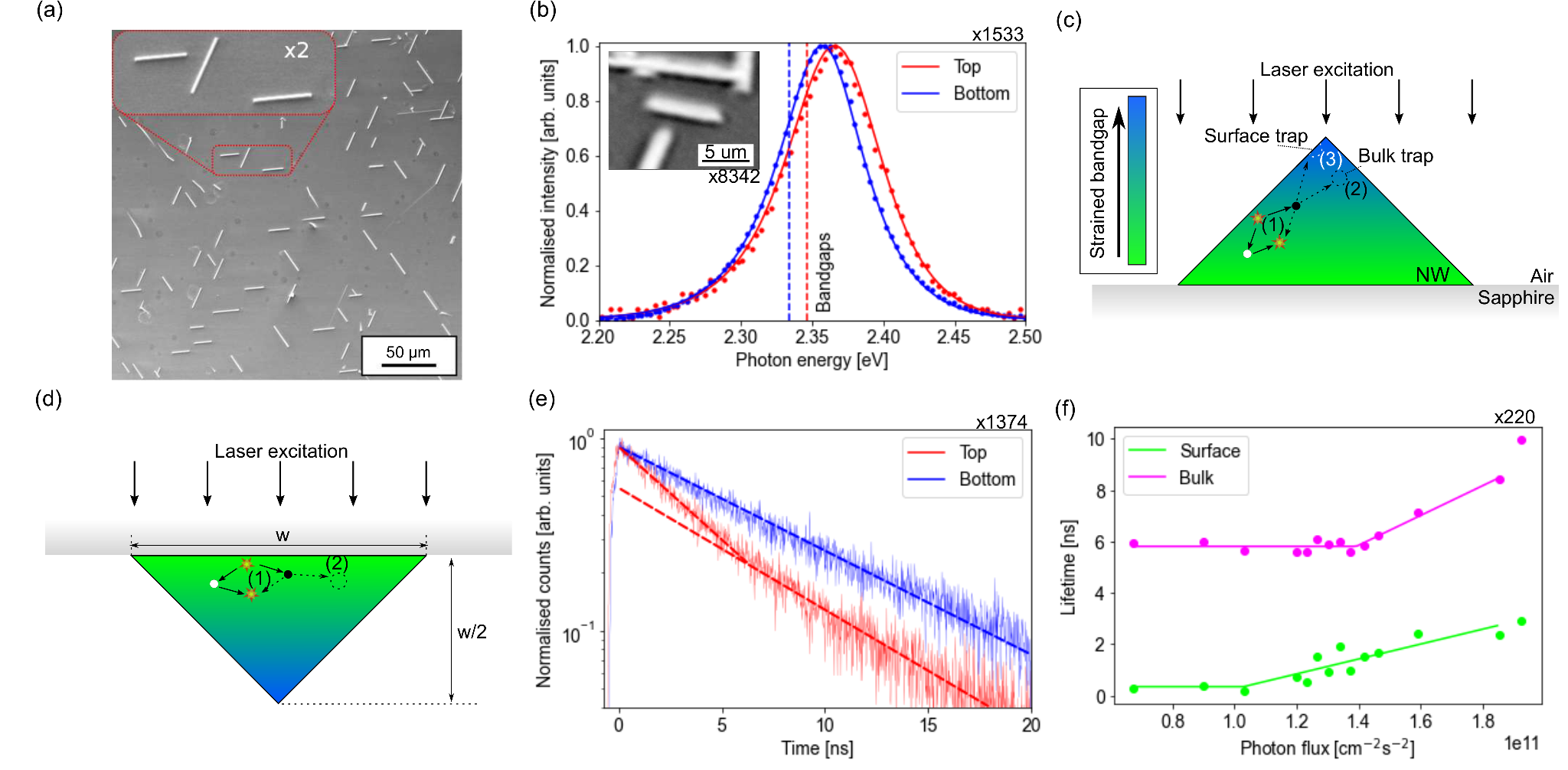}
    \caption{Single shot measurements of a NW subset. (a) A SEM image of a small population of CsPbBr\textsubscript{3} NWs showing the inter-NW variation of the length and width, along with the intra-NW uniformity of the width. Inset is a magnified image of 3 NWs. (b) PL spectra (dots) for top and bottom excitation of a single NW and the fit curve (solid lines) using a hybrid Urbach-Boltzmann given by Equation S3 in the S.I. An optical microscopy image of the same NW (inset). (c) A schematic of the NW cross section when laser excitation occurs from the top interface. Carriers are generated at the air/NW interface and are either trapped at the surface or diffuse throughout the coloured region before radiatively recombining or being trapped in the bulk. The bandgap varies throughout the cross section due to tensile strain at the NW/sapphire interface, resulting in a reduced bandgap at the base. (d) A NW schematic when exciting from the bottom, where carrier recombination is less sensitive to effects at the air/NW interface.(e) TCSPC data for top and bottom excitation of the same NW as (b). (f) The power dependence of the surface and bulk recombination lifetimes for a single NW. This data has been fit with a threshold model given by Equation S5 in the S.I. }
    \label{fig:1}
\end{figure*}

Microscopy imaging and spectroscopy was performed separately on more than 8000 NWs, an example of a single NW is shown in Figure~1b (inset). It takes approximately \SI{0.1}{\second} to locate a single NW and capture an image. PL spectra of this NW are also shown in Figure~1b: these spectra take approximately \SI{1}{\second} to acquire (including the time to move to the NW).

These spectra were fit with an emission model to extract the average PL bandgap (discussed in the methods section). The bandgap differs by \SI{13}{\milli\electronvolt} when exciting from the top and bottom, which are calculated to be \SI{2.348(1)}{\electronvolt} and \SI{2.335(1)}{\electronvolt}, respectively, provided with their standard errors (SE). This shift results from a lattice mismatch at the NW/substrate interface, leading to a reduction in the bandgap due to tensile strain at this interface~\cite{Oksenberg2018Surface-GuidedResponse}. The tensile strain relaxes with distance from the interface~\cite{Oksenberg2020}, therefore carriers recombining further from the NW/substrate interface emit higher energy photons: this is schematically illustrated in Figure 1c,d. When exciting from the top, the volume of material that the photocarriers can access spans a large range of strain environments, due to the variation in proximity to the lower interface. Conversely, when exciting from the bottom, the proximity to the lower interface is comparatively uniform and so the strain environments sampled are also uniform. This difference results in a larger degree of spectral inhomogeneity, which is shown in Figure S4a. in the S.I. These results demonstrate that the carrier diffusion length is smaller than the NW size, resulting in a non-uniform carrier distribution. 

TCSPC decays taken from the same NW are shown in Figure~1e, illuminating from the top and bottom. Each TCSPC measurement can take up to \SI{10}{\second} to acquire. When exciting from the bottom the decay is mono-exponential with a lifetime, and SE, of \SI{8.2(4)}{\nano\second}. Exciting from the top results in a bi-exponential decay. In this experiment, the peak carrier density is kept low (around \SI{1E16}{\per\centi\meter\cubed}) to avoid bi-molecular effects, and so the shape arises from two separate mono-exponential processes with lifetimes, and SEs, of \SI{3(2)}{\nano\second} and \SI{8(2)}{\nano\second}. 

The fast decay is only observed when exciting from the top, and so it is associated with carrier dynamics at the air/wire surface ~\cite{Yang2015}. Similar bi-exponential results are observed from PL measurements on CsPbBr\textsubscript{3} films~\cite{Hua2020}, and since the slow decay is observed in both measurements it is tentatively ascribed to carrier dynamics in the wire bulk. These assignments are supported by carrier diffusion models, shown in Figure S2 in the S.I., which demonstrate that the majority of carriers recombine in the bulk volume of the NW. These results also suggest that the bottom interface may be passivated.

It has previously been established that non-radiative carrier traps can dominate the carrier recombination in CsPbBr\textsubscript{3} materials~\cite{Yuan2018a}. The carrier densities used in this study are in the regime where these traps may be saturating~\cite{Jiang2021}, therefore this saturation process can be studied by varying the excitation power. An example of these power-dependent results from a single NW are shown in Figure~1f. Both lifetimes are constant at low excitation powers, and increase above a threshold due to saturation effects. This observation is compatible with previous studies which utilise higher carrier densities, and therefore observe longer carrier lifetimes to those reported here~\cite{Jiang2019,Oksenberg2021}. The threshold and the rate of change of the lifetime differ for the bulk and the surface - which likely reflects differences in the carrier traps in these regions.

Top/bottom single-element measurements provide unambiguous insight into surface effects, but provide no statistical strength or insight into correlation between parameters. Therefore, the full dataset can be used to exploit the variation in width to extract the link between NW geometry and dynamics. The NW widths were measured used optical microscopy and calibrated using SEM: this resulted in the range of widths of NWs studied to be between \SI{1.0}{\micro\meter} and \SI{1.4}{\micro\meter} (more statistics are provided in Figure S1 the S.I.). 

A total of 8342 NWs were imaged, taking a duration of 14 minutes to image the entire NW population. A subset of these NWs were studied for each measurement. The PL spectra were determined when exciting from the top and bottom for a subset of 1533 NWs, taking 26 minutes for each experiment. The PL lifetimes were measured for a subset of 3744 NWs from the top and 1737 from the bottom, taking approximately 5 hours and 4 hours respectively. In total, it took approximately 11 hours to record the 15576 individual measurements, with the vast majority of this time taken up by recording the TCSPC decays.

The correlation between PL-determined bandgap and NW width is shown in Figure~2a: in all cases the bottom illumination is redshifted relative to the top illumination. This energy shift depends on the tensile strain at the bottom interface, and the recombination positions of the carriers.  As the NW width is reduced, the bandgap blueshifts in both measurements. This has been previously attributed to lattice rotation effects, which increase the bandgap due to changes in the lattice bond angles~\cite{Oksenberg2020}. In a single NW, the lattice rotation is uniform, and so the overall effect is that the tensile strain relaxes and the bandgap increases with distance from the NW base, as shown in Figure~1e. The rotations increase in magnitude for narrower NWs~\cite{Oksenberg2018Surface-GuidedResponse}, resulting in larger bandgaps and energy shifts of up to \SI{60}{\milli\electronvolt} for NWs of \SI{1}{\micro\meter} width. There is a good correlation between our geometrical measurements and a previously relationship between geometry and emission energy~\cite{Oksenberg2018Surface-GuidedResponse}.

The lifetimes for top and bottom illumination are shown in Figure~2b,c respectively. The excitation-power dependence of the dynamics for a subset of 220 randomly chosen wires was also studied, and is presented in Figure S4 in the S.I. For the vast majority of NWs, bulk and surface recombination lifetimes are observed when exciting from the top, and only the bulk lifetime is observed from the bottom. As the NW width increases, the bulk lifetime becomes longer and the surface lifetime becomes shorter. This relationship between lifetime and NW width was investigated further with a model for carrier recombination.

\begin{figure*}
    \centering
    \includegraphics[width = \linewidth]{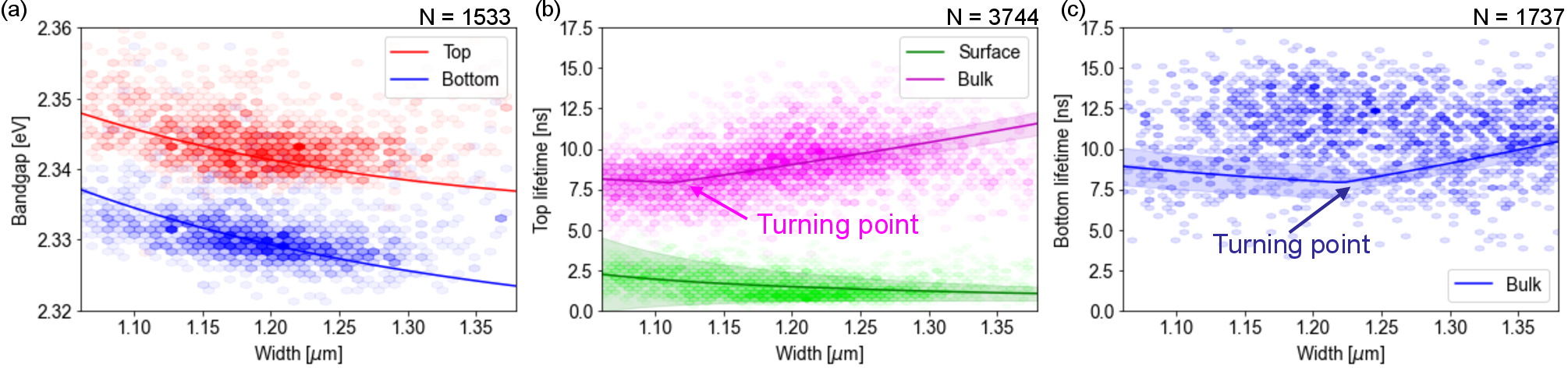}
    \caption{High-throughput optical results from the NW population, the shaded regions represent the SE on the model fit and \(N\) is the number of data-points in each data-set. (a) The bandgap obtained from fitting the PL spectra from 1533 NWs, correlated with the NW width. Data is shown when exciting from the top and bottom. The solid lines are fits of equations in the methods section, to the data. (b) The lifetimes, measured with excitation from the top, for the surface (fast) and the bulk (slow) recombination correlated with the width of 3744 NWs. The lines are fits of equations in the methods section to the data. The wide transparent area represents the 1$\sigma$ uncertainty of the model fit. (c) The bulk lifetimes, measured from the bottom as a function of the width of 1737 NWs. Turning points in the rates and carrier densities occur due to the onset of carrier trap saturation. The errors on the lifetime data have median values of 0.3, 0.5 and \SI{0.04}{\nano\second} for the top bulk and surface and bottom respectively: these numbers give reduced \(\chi^2\) values of 1.5, 0.1 and 5.8.}
    \label{fig:3}
\end{figure*}

\subsection{Model results}

To understand the impact of NW width on carrier dynamics, a self consistent model of diffusion and recombination was considered to explain the ensemble. Due to a small optical absorption length at the excitation wavelength (approximately \SI{100}{\nano\meter}~\cite{Maes2018}), carriers are generated close to the air/NW surface or the NW/substrate interface for top and bottom illumination respectively. These carriers can then diffuse throughout the NW and may recombine radiatively: alternatively the carriers may be trapped and recombine non-radiatively at the top interface or in the bulk~\cite{Jiang2019,Liu2021}. Transmission electron microscopy measurements do not observe the formation of any strain-related defects at the bottom interface~\cite{Oksenberg2020}, and the TCSPC measurements also show no evidence of an additional recombination mechanism.  For this reason, the bottom interface is treated like bulk material in the model. The model makes no assumptions regarding the nature of the carriers that dictate this behaviour, and also no assumptions about the nature of the carrier traps. Candidates for these traps include dangling bonds at the surface~\cite{Lorenzon2019}, Pb vacancies and Br interstitials~\cite{Yuan2018a}.

The model relies on the observation that carrier diffusion occurs on a timescale faster than recombination~\cite{Herz2017}. A schematic of this process is shown in Figure~1c,d. Carrier diffusion is accounted for using a time-dependent Monte-Carlo model that is shown in Figure S2 in the S.I. This produces a distribution of carrier recombination locations in the NW that is used to calculate the carrier densities and dynamics, as described in the methods section. Table~1 provides a summary of the most important extracted optoelectronic parameters from this model: all uncertainties quoted in this section are the SE extracted from the model.

\begin{table}
\centering
\begin{tabular}{llcc}
\hline
Parameter & Unit & Value                  &  literature \\ \hline
Bulk trap density: $N_V$ & [$10^{16}$\si{\per\centi\meter\cubed}]       & \(8.6 \pm 0.4\) &  $<$ 15\textsuperscript{(a)}~\cite{Huang2019a}\\
Surface trap density: \(N_S\) & [$10^{16}$\si{\per\centi\meter\cubed}]        & \(7.1 \pm 0.3\) &  $<$ 12~\cite{Ni2021}\\
Unstrained bandgap: \(E_g\)  &[\si{\electronvolt}]       & \(2.4 \pm 0.1\)                   &  2.36~\cite{Mannino2021}\\
Diffusion length: \(L_D\)  &[\si{\micro\meter}]     & \(0.25 \pm 0.02\)                   &  9.2\textsuperscript{(b)}~\cite{Yettapu2016}\\
IQE (top) &[\si{\percent}]     & \(0.7 \pm 0.1\)                   &  -\\
Carrier mobility: \(\mu\) & [\si{\centi\meter\squared\per\volt\per\second}]  & \(0.8 \pm 0.1\)                   & 35\textsuperscript{(c)}~\cite{Oksenberg2021}\\
\hline
\end{tabular}
\caption{Optoelectronic parameters derived from the model, with their SEs, compared with literature values from single-shot studies (where available). (a) Reports for spin coated CsPbBr\textsubscript{3} LEDs. (b) Measured for CsPbBr\textsubscript{3} single nano-crystals. (c) Reports for carrier diffusion along the long axis of CsPbBr\textsubscript{3} NWs.}
\label{tab:results}
\end{table}

The PL bandgap varies with NW width and carrier recombination location in the wire. The model allows us to separate the effects of tensile strain and lattice rotations from the data in Figure~2a, using the equations defined in the methods section. As the NW width increases, the lattice rotation effects reduce and the bandgap redshifts (equally for top and bottom excitation). This lattice rotation effect accounts for a bandgap shift of \SI{12}{\milli\electronvolt} across the NW ensemble, which is comparable to values previously measured using photoluminescence on a small number of NWs of this size~\cite{Oksenberg2018Surface-GuidedResponse}. Increasing the width also increases the average separation between carriers when exciting from the top and bottom - this causes a larger shift between the two bandgap measurements. This is a comparatively small effect, accounting for an additional \SI{2}{\milli\electronvolt} bandgap shift across the ensemble. The multi-modal fit determines the unstrained bandgap to be \SI{2.4(1)}{\electronvolt}, which is consistent with literature values~\cite{Mannino2021}. It is notable that this result cannot be obtained from a single-wire measurement because none of the NWs in this study have unstrained emission.

The calculated ambipolar diffusion length is \SI{0.25(2)}{\micro\meter}, which is smaller than best-in-class literature values~\cite{Stranks2013,Yettapu2016}. This is smaller than the NW cross section, validating the assumption that diffusion will be an important in carrier recombination behaviour, and that the carrier distribution is non-uniform. Additionally, the carrier mobility is \SI{0.8(1)}{\centi\meter\squared\per\volt\per\second}, comparable with catalyzed vapor-liquid-solid grown NWs~\cite{Meng2019}. However, there is a wide spread of mobilities reported in the literature due to variation in sample quality and experimental approaches~\cite{Yettapu2016,Wolf2018}. Our mobility is smaller than previous reports on NWs grown using the same method, studying diffusion along the NWs~\cite{Oksenberg2021}. This discrepancy may partially be due to differences in excitation conditions, and carrier densities, as well as the nature of the high-throughput study, which accounts for the entire NW population, including those of poorer quality. There is also the possibility of mobility anisotropy along and across the NW cross section, due to strain or compositional variation in each direction~\cite{Frohna2021}.

The model fits to the lifetime measurements are shown in Figure.~2b,c. The model accounts for the trends in lifetime with NW width when exciting from the top. The larger relative uncertainties on the fast lifetimes are reflective of the increased uncertainties on the lifetimes measured from each NW. The bottom lifetime data has a greater spread than is accounted for in the model; this may be experimentally driven, arising from challenges in focusing the microscope objective through the sapphire substrate. This data still provides a useful constraint on the holistic model. 

The observed lifetime trends are primarily influenced by trap saturation effects, which allow the density of traps to be calculated. At the surface this is \SI{7.1(3)E16}{\per\centi\meter\cubed}, which comparable to the trap density in the bulk, \SI{8.6(4)E16}{\per\centi\meter\cubed}. However, the surface traps lie within a small depth, leading to a high areal density of \SI{4.3(7)E10}{\per\centi\meter\squared} and increasing the surface recombination rates. Accounting for the traps in the surface and bulk regions, the average number of traps in a single NW is \SI{6.3(3)E8}{}. For comparison, an excitation pulse with \SI{2E11}{photons}.cm\textsuperscript{-2} generates \SI{2.8(1)E8} carriers in a NW, averaged over the NW widths considered in this study. The excitation conditions in this experiment are therefore sufficient to study trap saturation effects and the trap densities lie in the range of literature values~\cite{Huang2019a,Ni2021}. 

The holistic approach, which analyzes and correlates all the measured properties for each single element in the ensemble, has therefore determined material parameters that are compatible with literature values obtained from single-shot measurements (where available), and has established values with statistical confidence that represent the entire NW population.

\begin{figure*}
    \centering
    \includegraphics[width = \linewidth]{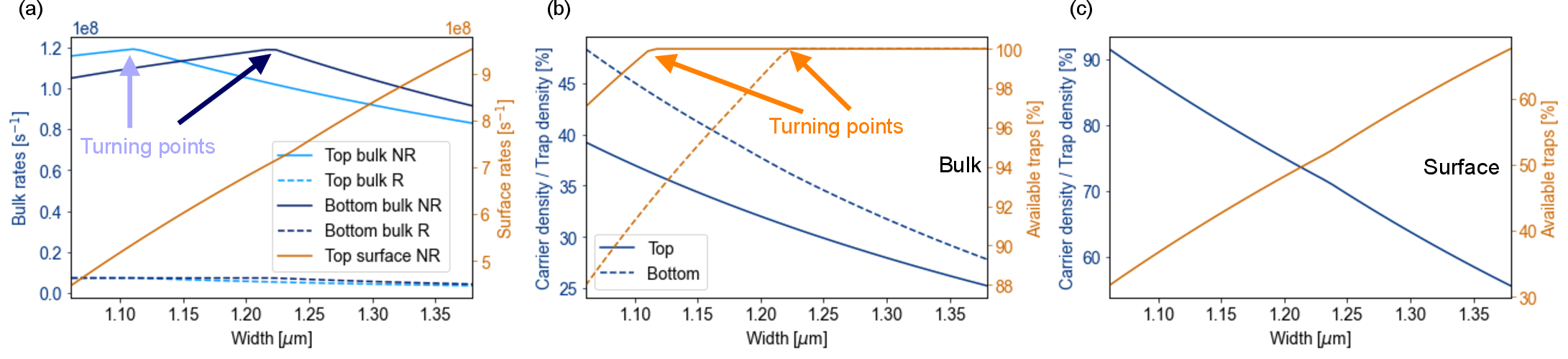}
    \caption{Derived parameters from the global model fit for NWs with different widths. (a) Calculated radiative (R) and non-radiative (NR) rates. Turning points in the rates and carrier densities occur due to the onset of carrier trap saturation. (b) Carrier and unoccupied trap densities in the bulk when exciting from the top and bottom, normalised to the trap density. (c) Carrier and unoccupied trap densities at the surface when exciting from the top, normalised to the trap density.}
    \label{fig:4}
\end{figure*}

The model can describe the variation in the carrier recombination rates for different NW widths, as shown in Figure~3a. The mean non-radiative rate at the surface is \SI{7(1)E8}{\per\second}, and dominates over the other processes. This is compatible with conclusions from other studies~\cite{Liu2021}. The mean bulk non-radiative rate is \SI{1.0(1)E8}{\per\second}, which is much faster than the mean radiative rate of \SI{5(1)E6}{\per\second}. Therefore, non-radiative effects dominate the carrier dynamics, as predicted by previous studies~\cite{Jiang2019}.

The lifetime variation with NW width can be explained by considering the effect of trap and carrier densities. As shown in Figure~3b,c, as the wire width increases the carrier density reduces at the surface and in the bulk. As the excitation conditions are constant for each measurement, this is due to an increased wire volume and surface. This causes a reduction in the radiative rate with increasing wire width in Figure~3a. 

The non-radiative rate is influenced by the carrier density and the effective trap density, given by the trap density minus the saturated trap density. As the carrier density increases above a threshold, the effective trap density will reduce due to saturation. This can be seen for the narrowest NWs in Figure~3b. The overall change in the non-radiative rate will be a combination of these effects.

Figure~3a shows that the bulk non-radiative rate from the top increases with width up to \SI{1.12}{\micro\meter}. This occurs because the trap density is limiting the non-radiative rate, and there is an increase in the available traps. The trap saturation can be as high as \SI{13}{\percent} in these NWs. For wider NWs, there is effectively no trap saturation - and so the available traps remains constant at \SI{100}{\percent}. The non-radiative rate is now limited by carrier density, and so drops with increasing width. This results in the turning points in the model, and the lifetime increase observed in Figure~2b. As the turning point is related to trap saturation, it is also dependent upon the excitation fluence used in the experiment: for example, a higher fluence would cause a greater degree of trap saturation, and shift the turning point to wider NWs. At the surface, the carrier density is sufficient to saturate a portion of the surface traps for all NW widths, as shown in Figure~3c, leaving between \SI{35}{\percent} and \SI{70}{\percent} of traps available for capture. As a result, the non-radiative rate increases with wire width, and the lifetime in Figure~2b drops.

When exciting from the bottom, the carrier density in the bulk is higher due to stronger coupling of the incoming light, as demonstrated in Figure S1 in the S.I. The higher carrier densities mean that the radiative rate is, on average, \SI{25}{\percent} faster at the bottom. However, this change is insignificant when compared to the non-radiative rates. Additionally, more bulk traps are saturated for the same NW width. This shifts the turning point in the non-radiative rate to wider NWs. The higher carrier density partially balances out with the trap saturation and so the bulk non-radiative rate from the bottom is comparable to that from the top, demonstrated in Figure~3a. As shown in Figure~2c, this means that the fit to the bottom lifetime is of a similar shape.

The model can also be used to predict opto-electronic behaviour of the NWs with statistical confidence. Equation~8 in the methods section was used to calculate the average recombination IQE when exciting from the top, which was found to be \SI{0.7(1)}{\percent}. This result was independently verified using temperature dependent PL to estimate the IQE of a small number of NWs. The mean room temperature IQE was measured to be \SI{1.1(3)}{\percent}, which is consistent with the model calculation. Since the NWs have a significant variation in geometric parameters and have a strongly non-uniform coverage across the substrate, these individual single-NW IQE measurements are more meaningful that an ensemble PL-quantum yield measurement. More details of the temperature dependent PL, along with the variation of the IQE with width are shown in Figure S4 in the S.I.

The IQE can be increased by exciting the carriers at the bottom of the wire, which will reduce the number of carriers that diffuse to the surface. This will reduce the overall rate of non-radiative recombination: the model predicts that the IQE will increase to \SI{5.8(4)}{\percent}. This prediction was again confirmed using temperature dependent PL, which is shown in Figure S6 of the S.I. to be \SI{7(1)}{\percent}.

\section{Conclusion}
This paper reports on a  multi-modal high-throughput spectroscopy technique that harnesses the variation in geometry amongst a nano-material population to extract optoelectronic properties and performance. This is achieved through measurements of optical images, PL spectra and PL decays of the same nano-materials when illuminating from two different directions. The analysis of global trends across the population enables the extraction of a multitude of important optoelectronic parameters with statistical confidence that are comparable to separately determined values in the literature.

The technique has been applied to CsPbBr\textsubscript{3} NWs, only requiring an independent measurement of the cross-section shape and the refractive index. The PL bandgap increases with reduced NW thickness due to increasing bond rotation effects, and reduces at the wire/substrate interface due to tensile strain. Non-radiative recombination at defects at the air/NW surface is the dominant carrier recombination process, whilst non-radiative recombination at bulk defects and radiative recombination are also observed. As the NW width increases, the bulk recombination lifetime increases, and the the surface recombination lifetime decreases. A self-consistent model of carrier diffusion and recombination was developed to explain these trends and showed that they are due to changes in carrier density, and trap saturation, with NW width. Furthermore, the model extracts the carrier diffusion length, mobility, trap densities and bandgap from the multi-modal data - giving results which are compatible with the literature. Crucially, the technique is able to accurately predict the IQE of carrier recombination, and how this changes when illuminating from different directions. As minimal \latin{a-priori} information is required, this approach provides a data-driven methodology to explore nanomaterial systems.

\section{Experimental Section}
\subsection{Sample growth}
The surface-guided NWs were synthesized via a vapour-phase method on \textit{c}-plane sapphire substrates. A spectroscopic hotplate (Linkam THMS600) is used, with a round stage size of \(\approx\) \SI{2}{\centi\meter} radius, over which a silicon wafer is placed as a holder for chunks of molten CsBr and PbBr\textsubscript{2} powders (both 99.999~\%, purchased from Sigma-Aldrich). The sample was placed face down over an aluminium ring spacer, such that the distance from the chunks is \(\approx\) \SI{500}{\micro\meter}. The stage is then heated to \SI{420}{\degreeCelsius} with a rate of \SI{30}{\degreeCelsius\per\minute}, and kept at the NW growth temperature for 5-10 min, after which the stage is cooled down to room temperature at a rate of \(\approx\) \SI{100}{\degreeCelsius\per\minute}.

\subsection{Scanning electron microscopy}
SEM was performed using an Quanta250 FEG microscope to acquire images of the NWs. The images were obtained using a secondary electron detector at 11.3 mm working distance and 5 keV acceleration voltage. Images acquired at 500x magnification show $\approx$ 120 NWs per image. 

\subsection{High throughput spectroscopy}
NWs on their substrates were placed into a home-built microscope with a quasi-confocal setup for optical excitation and light collection. A motorized xyz translation system was used for brightfield microscopy with an optical resolution of approximately \SI{1}{\micro\meter}. A machine vision camera was used to identify NWs by their geometrical properties (width, length, orientation etc.). A total of 8342 NWs were identified using this method, taking approximately \SI{0.1}{\second} per NW. The NW widths in this study were close to the resolution of the microscope, and so these measurements were calibrated using SEM and AFM, as shown in Figure S1 in the S.I.

Room temperature photoluminescence spectroscopy was performed using a \SI{405}{\nano\meter} wavelength pulsed laser, with a \SI{50}{\nano\second} pulse period and \SI{80}{\pico\second} pulse duration, to excite the NWs in their centre. The laser spot was focused to an elliptical spot size of \SI{8}{\micro\meter} by \SI{21}{\micro\meter} and the photon flux at the NW was estimated to be \SI{2E12}{\per\centi\meter\squared\per pulse}. A longpass filter was used to remove the excitation light and the luminescence was focused onto a fibre connected to an Ocean Optics QE65000 spectrometer with a spectral resolution of \SI{0.3}{\nano\meter}. This experiment was automated to sequentially measure the PL of the full NW population, taking approximately \SI{1}{\second} per NW, including movement time. A schematic of this system is shown in Figure S2 in the S.I.

The PL time decays of NWs were measured by routing the output optical fibre to single photon avalanche diodes connected to a Picoharp HydraHarp400 TCSPC system. The combined experimental setup has a time resolution of \SI{200}{\pico\second}, and it takes between 1 and \SI{10}{\second} to measure a single wire, including movement time. The excitation power was varied using a motorised ND filter wheel. Power dependent TCSPC was measured for a subset of 220 NWs.

PL measurements were obtained by exciting NWs from the top and from the bottom. This was achieved by flipping the sample substrates in the microscope and repeating the machine vision imaging to identify another set of NWs, and repeating the high throughput spectroscopy. The NWs that were present in both populations were identified using an approach similar to reference ~\cite{Lang2010}. This resulted in 1533 NWs where both sets of measurements were performed.

\subsection{PL fitting}
The PL spectra were fit with a model assuming that photogenerated carriers rapidly cool to the conduction and valence band edges and occupy a thermal distribution in a 3D density of states. This density of states was modified to include an Urbach tail at low energies, and was convoluted with a Gaussian to account for inhomogeneity in the NW and the system resolution. More details of the fitting can be found in Figure S3 in the S.I.

\subsection{TCSPC fitting}
All PL decays were fit with both a mono-exponential and a bi-exponential fit. The reduced chi-squared of these fits was assessed to determine if the measurements contained one or two decays. The same procedure was applied to the power dependence of the decays to obtain how the lifetimes vary with excitation power. A linear threshold model was fit to the lifetime vs excitation power data to extract the threshold power and the rate of lifetime change above the threshold, to constrain the recombination model. More details and examples of the TCSPC fitting can be found in Figure S3 in the S.I.

\subsection{Recombination model}
A quasi-steady state model for photon absorption, carrier diffusion and recombination was developed to fit the PL and TCSPC data. This includes the PL bandgap data and the PL lifetimes from Figure~2. The model is also constrained by the power dependence of the PL lifetimes and the integral of the PL decays, which are shown Figure S4 of the S.I. The model only requires prior knowledge of the cross-sectional shape and the refractive index of the material.

This model accounts for geometrical factors in the NW cross section, which differ when they are excited from the top or the bottom, as shown in Figure~1. COMSOL modelling is used to determine the average fraction of incident light which is absorbed in the NW, as shown in Figure S1 in the S.I., $\langle abs \rangle$. The number of carriers generated in the NW, $N_{\rm{0}}$ is given by Equation~1:
\begin{equation}
\label{equ:number}
N_{\rm{0}} = \langle abs \rangle\ \langle w_{\rm{spot}} \rangle\ n_{\rm{flux}}
\end{equation}
where \(\langle w_{\rm{spot}} \rangle\) is the average width of the excitation spot, \(n_{\rm{flux}}\) is the photon flux of the excitation spot and \(N_{\rm{0}}\) is in units of \si{cm^{-1}}, which is low enough to ignore Auger recombination. As the absorption length is \(\approx\)\SI{100}{\nano\meter}~\cite{Maes2018}, the calculations assume that these carriers are generated at the surface.

Carrier diffusion occurs on sub-ns timescales~\cite{Herz2017}: the model therefore assumes that carriers diffuse throughout the NW cross section before recombining, with a characteristic diffusion length \(L_{\rm{D}}\). This diffusion is modelled using a Monte-Carlo simulation of nearest neighbour hopping, the details of which are shown in Figure S2 in the S.I. Carriers that stop diffusing within a certain distance from either interface, \(L_{\rm{surface}}\), are trapped at that interface: this is used to calculate an occupation factor \(B\), which represents the proportion of carriers at the top surface, the bottom interface and in the bulk volume. The volume (\(V\) [\si{\centi\meter\squared}]) of each of these regions is geometrically calculated and used in Equation~2 to find the photogenerated carrier densities (\(n\) [\si{\per\centi\meter\cubed}]):
\begin{equation} 
\label{equ:densities}
  n= \frac{\langle abs \rangle\ \langle w_{\rm{spot}} \rangle\ n_{\rm{flux}}}{V} B
\end{equation}
Equation~1,2 apply to illumination from the top and bottom, with different values of \(\langle abs \rangle\) and \(B\). At the top surface, non-radiative recombination dominates~\cite{Jiang2019,Liu2021} - it is therefore assumed that all of the carriers at the interfaces are trapped. In the bulk volume, carriers are either free or trapped. The carrier densities are calculated using Equation~3,4:
\begin{align} 
\label{equ:carr_dens}
n_{\rm{S}} &= n_{\rm{S,trapped}}, \\
n_{\rm{V}} &= n_{\rm{V,free}} + n_{\rm{V,trapped}}.
\end{align}
The free carriers undergo radiative recombination with a rate proportional to the density of free carriers squared, and is assumed to be negligible at the top surface. Non-radiative recombination is mediated by the effective trapping rate, which is proportional to the density of unoccupied traps, \(n_{\rm{t,eff}}\) and the density of free carriers. The total recombination rates are given by Equation~5,6:
\begin{align} 
\label{equ:rates}
{\frac{1}{\tau_{\rm{S}}}} &= {k_{\rm{tS}} n_{\rm{tS,eff}} n_{\rm{S,free}}},\\
{\frac{1}{\tau_{\rm{V}}}} &= {k_{\rm{rV}} n_{\rm{V,free}}^2 + k_{\rm{tV}} n_{\rm{tV,eff}}} n_{\rm{V,free}}.
\end{align}
where \(k\) are the rate constants and \(\tau\) is the recombination time measured by TCSPC. Power dependent TCSPC shows how these lifetimes increase when the excitation power increases above a threshold. This allows the model to constrain the trap density in each region using Equations S6-S10 in the S.I.

The bandgap of the NW is redshifted due to tensile strain and is blueshifted by lattice rotation effects~\cite{Oksenberg2020}. These rotation effects are uniform throughout an individual NW, and are stronger in narrower NWs~\cite{Oksenberg2020}. The tensile strain relaxes with distance from the NW base such that it is unstrained at the top. Therefore, carriers generated by top and bottom illumination will experience the same lattice rotation effects, but different degrees of strain, as illustrated in Figure~1c. 

The average bandgap of carriers is modelled using the Monte-Carlo diffusion model. The position of all the carriers in the ensemble is used to find the average distance from the bottom interface, \(y_{\rm{av}}\). This distance is then included in an empirical equation that is fit to the data in Figure~2a. A simplification is made to the case in reference~\cite{Oksenberg2020}, assuming that the tensile strain relaxes linearly with distance from the bottom interface - and is unstrained at the top apex. An exponential variation of energy with NW width is assumed, with constant \(L_{\rm{rot}}\), due to lattice rotation effects. The bandgap is given by Equation~7: 
\begin{equation}
\label{equ:bandgaps}
E_{\rm{g}} =  E_{\rm{0}} - T \frac{2y_{\rm{av}}}{w} + A \text{exp}\left(\frac{-w}{L_{\rm{rot}}}\right)
\end{equation}
where \(E_{\rm{0}}\) is the unstrained bandgap of the NW, \(A\) is a scaling factor, \(w\) is the NW width and \(T\) is the energy shift due to tensile strain at the base. 

This model produces 13 fitting parameters in total. A detailed description of these parameters is provided in the SI. The fitting parameters can be used to derive values for the IQE, defined as the total radiative recombination rate, divided by the total decay rate, this is given by Equation~8:
\begin{equation}
\label{equ:IQE}
IQE = \frac{k_{\rm{rV}} n_{\rm{V,free}}^2}{k_{\rm{rV}} n_{\rm{V,free}}^2 + k_{\rm{tV}} n_{\rm{tV,eff}} n_{\rm{V,free}} + k_{\rm{tS}} n_{\rm{tS,eff}} n_{\rm{S,free}}}.
\end{equation}
Note that, for illumination from the bottom, \(n_{\rm{S,free}}=0\). The carrier mobility, \(\mu\), can also be calculated using Equation~9:
\begin{align}
\label{equ:mobility}
{\mu = \frac{e L_{\rm{D}}^2}{kT \tau_{\rm{V}}}.}
\end{align}

\subsection{Additional techniques}
To supplement the high throughput measurements, a subset of NWs were studied using AFM. Temperature dependent PL was also used to estimate the IQE. The coupling of the excitation beam into the NWs was studied theoretically and carrier diffusion was simulated using Monte-Carlo methods. Details of these techniques are provided in the S.I.

\begin{acknowledgement}
This was work funded by UKRI under grant MR/T021519/1, the Israel Science Foundation under grant 2444/19 and the Minerva Center for Self-repairing systems for energy and sustainability. Research data supporting this publication will be made available at DOI: 10.48420/19746019, and the code to perform the modelling will be made available at DOI: 10.48420/19746757, and on github:\\ \url{https://github.com/p-parkinson/holistic_determination_CsPbBr3_nanowires}.

CRediT author statement: \textbf{Stephen Church}: Formal analysis, Investigation, Methodology, Software, Visualisation, Writing - original draft. \textbf{Hoyeon Choi}: Investigation. \textbf{Nawal Al-Amairi}: Investigation, Writing - review and editing. \textbf{Ruqaiya Al-Abri}: Investigation, Writing - review and editing. \textbf{Ella Sanders}: Resources, Writing - review and editing. \textbf{Eitan Oksenberg}: Writing - review and editing \textbf{Ernesto Joselevich}: Funding acquisition, Resources, Writing - review and editing. \textbf{Patrick Parkinson}: Conceptualization, Data curation, Funding acquisition, Methodology, Software, Supervision, Writing - review and editing. 

The authors thank Andras Botar for his assistance with data management.
\end{acknowledgement}

\begin{suppinfo}
AFM of a subset of the NWs compared with SEM and optical microscopy, temperature dependent PL to estimate the IQE and additional statistics from the NW population. Details of the models for PL and TCSPC fitting, theoretical analysis of light coupling into the NWs, Monte-Carlo simulations of carrier diffusion and a summary of the parameters in the holistic model.
\end{suppinfo}

\providecommand{\latin}[1]{#1}
\makeatletter
\providecommand{\doi}
  {\begingroup\let\do\@makeother\dospecials
  \catcode`\{=1 \catcode`\}=2 \doi@aux}
\providecommand{\doi@aux}[1]{\endgroup\texttt{#1}}
\makeatother
\providecommand*\mcitethebibliography{\thebibliography}
\csname @ifundefined\endcsname{endmcitethebibliography}
  {\let\endmcitethebibliography\endthebibliography}{}


\renewcommand{\thefigure}{S\arabic{figure}}
\renewcommand{\thetable}{S\arabic{table}}
\renewcommand{\theequation}{S\arabic{equation}}
\setcounter{figure}{0} 
\setcounter{equation}{0} 
\setcounter{table}{0} 

\newpage
\section{Supporting Information}
\subsection*{Scanning electron microscopy and atomic force microscopy}
\label{sec:COMSOL}
\begin{figure*}[h!]
    \centering
    \includegraphics[width=1\linewidth]{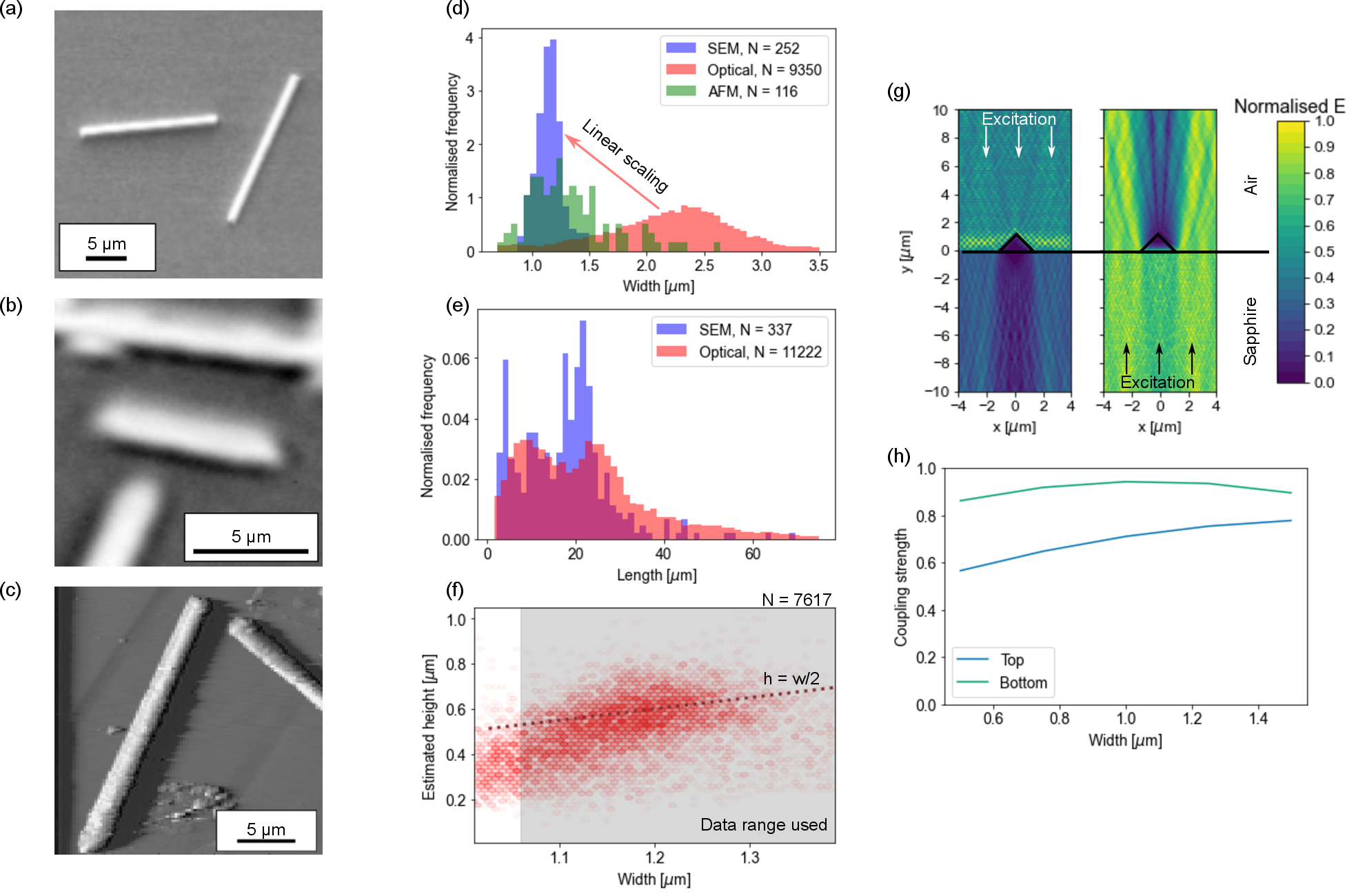}
    \caption{A summary of the geometrical characterisation and light coupling modelling of NWs. Images of CsPbBr\textsubscript{3} NWs taken using (a) SEM, (b) optical microscopy and (c) AFM. (d) A comparison of the width measurements for NWs using different techniques, 252 NWs were studied with SEM, 9350 using optical microscopy and 116 using AFM. (e) A comparison of the NW lengths using SEM and optical microscopy, 337 NWs were studied with SEM and 11222 using optical microscopy. (f) The correlation between the height, estimated from the PL peak energy, and the width for 7617 NWs. The shaded region represents the range of NWs considered in this study. The dotted line is a guide to the eye for h = w/2. (g) The normalised electric field distribution when exciting a NW from the top and bottom. (h) The coupling strength of light into NWs with different widths when exciting from the top and bottom.}
    \label{fig:COMSOL}
\end{figure*}
Scanning electron microscopy (SEM), optical microscopy and atomic force microscopy (AFM) were used to measure the lengths and widths of nanowires (NWs). Example images for these techniques are shown in Figure~S1a-c. The SEM and optical results are obtained using the systems described in the main paper. AFM was performed using a Nanosurf system in tapping mode with a Tap190Al-G probe. The maximum heights of the NWs were measured and the width was calculated by dividing the values by two, due to the triangular cross section~\cite{Oksenberg2020a}. Histograms of these results are shown in Figure~S1d,e.

The width distribution is comparable for SEM and AFM results, with median, and standard deviation (SD), values of \SI{1.2(3)}{\micro\meter} and \SI{1.2(4)}{\micro\meter} respectively. This provides evidence for the triangular cross sectional shape previously reported~\cite{Oksenberg2020a}. The NW widths for optical measurements are significantly larger, with median and SD values of \SI{2.2(6)}{\micro\meter}. These larger values are because the widths are of the same order as the resolution of the optical microscope. To account for this a linear scaling was applied to the optical widths in order to replicate the SEM histogram.

The NW length distribution is similar for SEM and optical results: the median and SD of the optical length was \SI{21(1)}{\micro\meter} and this was \SI{18(1)}{\micro\meter} in the SEM results. The distribution is bi-modal, with peaks at \SI{9}{\micro\meter} and \SI{24}{\micro\meter} in the optical measurements. The heights of the NWs were also estimated using a previously determined empirical relationship between emission energy when illuminating from the top and nanowire height~\cite{Oksenberg2018Surface-GuidedResponsea}. A polynomial was fit to this previously reported data and applied to the PL spectra of 7617 NWs. The results of this conversion are shown in Figure~S1f. As the NW width approaches the diffraction limit, the linear-scaling approximation breaks down due to an increase in the relative uncertainty in the width. We determined a validity range for our linear-scaling by performing a series of linear fits to the height-width relationship for different ranges of NW widths. We found that the h = w/2 relationship holds for NWs wider than \SI{1.06}{\micro\meter}, and thus we selected these wires for further experimentation. These NWs represent \SI{90}{\percent} of the total population demonstrate the consistency between results from previous studies on the same wires ~\cite{Oksenberg2018Surface-GuidedResponsea}, the AFM widths (calculated from the height measurements) and the width calibration procedure using SEM and optical microscopy. 

For narrow NWs, the estimated height, and SE, is \SI{0.15(10)}{\micro\meter} less than that expected from the $h = w/2$ relation. This may be because these wires lie close to the resolution limit of the optical microscope: the width of the NWs in this region may therefore be over-estimated and these wires were removed from any analysis.

\subsection*{Light coupling modelling}
\label{sec:COMSOL2}

The coupling of light into the NWs was calculated using the quadratic finite-element-method for different NW widths. A triangular cross section of an ideal NW was modelled in two dimensions with widths ranging from \SI{0.5}{\micro\meter} to \SI{1.5}{\micro\meter}. The NW was sandwiched between a sapphire substrate and the air. To account for the elliptical excitation spot, these calculations were repeated for modelling spaces with widths of \SI{8}{\micro\meter} and \SI{21}{\micro\meter}, each representative of an axis of the ellipse. The calculations were also repeated for top and bottom illumination. Example modelling geometries are shown in Figure~S1g. The real and imaginary refractive indices of the NW were defined as 2 and 0.32 respectively \cite{Maes2018a,Zhao2018} and the refractive index for the sapphire was 1.77~\cite{Malitson1972}. COMSOL multi-physics was used to perform the calculations, using the "electromagnetic wave, frequency domain (ewfd)" package to solve Maxwell's equation in three dimensions, given by Equation~S1:
\begin{equation}
 \label{eqn:model}
    \nabla\times(\mu_{\rm{r}}^{-1}\nabla\times\textbf{E}) - \omega^2/c_{\rm{o}}^2(\epsilon_{\rm{r}} - \frac{j\sigma}{\omega\epsilon_{\rm{o}}})\textbf{E} = 0,
\end{equation}
where \textit{\textbf{E}} is the electric field, $\sigma$ is the electrical conductivity, $\omega$ is the angular frequency, $\mu_r$ and $\epsilon_r$ are the relative permeability and permittivity respectively. The results were averaged for electric field polarisation in and out of the plane.

The simulation accounts for reflection and transmission of light at each interface, scattering and diffusion of light from the interfaces, and absorption of light in the NWs. An example calculated electric field distribution is shown in Figure~S1g. The proportion of the incident light intensity which is absorbed by the NW was calculated by considering the difference in intensity of the excitation and after the interaction. The intersection between a single NW and the excitation spot will vary depending on the NW orientation: to approximate this effect the result was averaged between the \SI{8}{\micro\meter} and \SI{21}{\micro\meter} configurations. To calculate the light coupling strength, the absorption was divided by the absorption of a block of perfectly absorbing material with the same dimensions as the NWs. The coupling strengths are shown in Figure~S1h. The bottom coupling also includes the reflectivity of the initial air/sapphire interface at the bottom of the sample, calculated to be 0.077. 

The coupling strength when illuminating from the bottom remains approximately unchanged with wire width, with a value and SD of \SI{0.91(3)}. The coupling strength is high due to smaller changes in refractive index at the two interfaces, reducing the Fresnel reflection at the interfaces. The coupling is constant because light is normal to a flat interface, which is unchanged by changing the width. However, for top illumination the coupling strength is 0.78 for wire widths close to \SI{1.5}{\micro\meter} and reduces for smaller widths. The reduced coupling is partially due to enhanced reflection from a larger change in refractive index. Additionally, scattering from the apex of the triangle increases for smaller NWs.

\subsection*{Carrier diffusion model}

\begin{figure*}
    \centering
    \includegraphics[width=1\linewidth]{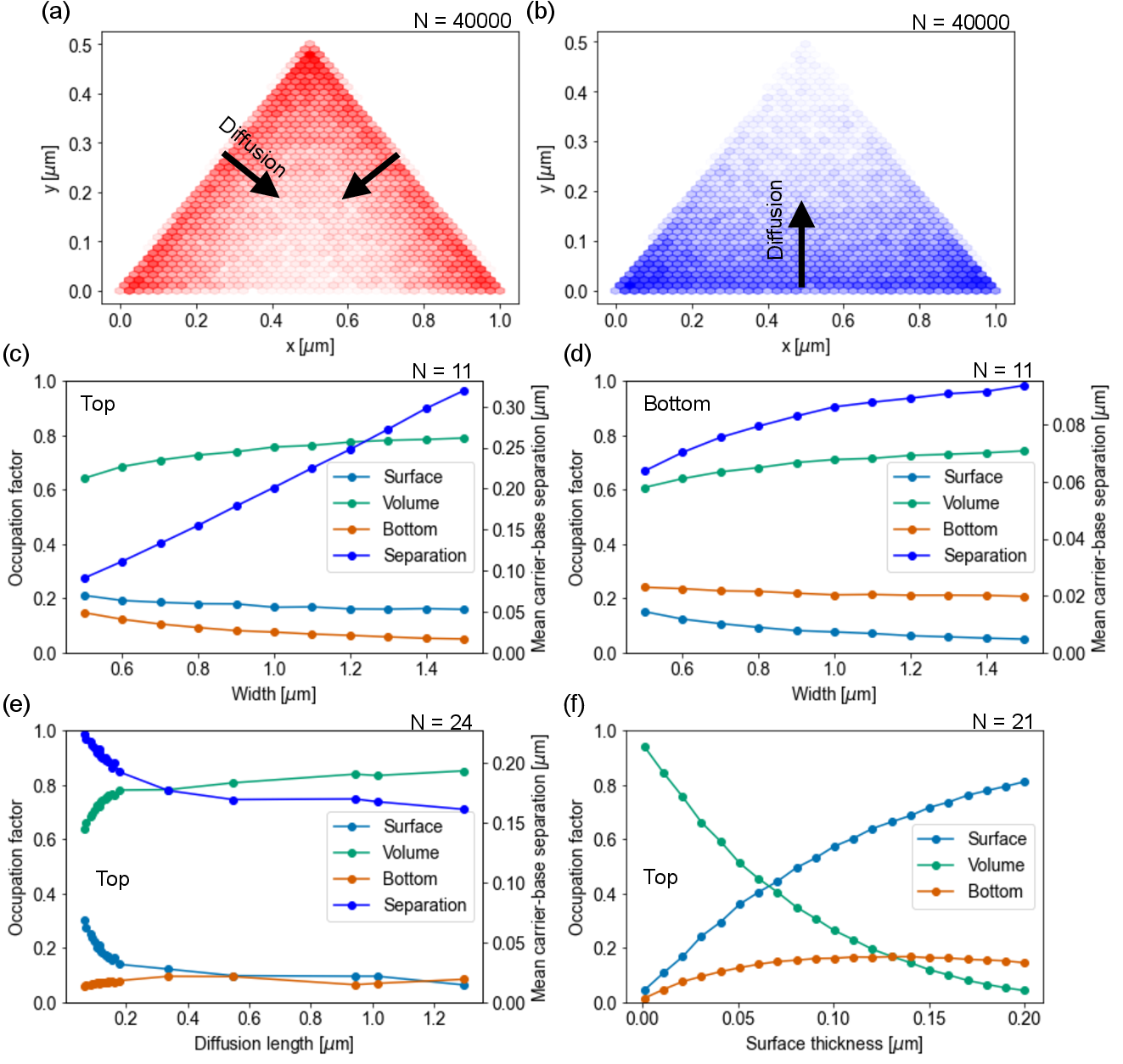}
    \caption{Summary of the results of the Monte-Carlo carrier diffusion model. (a) The distribution of carrier recombination positions when illuminating from the top of a NW with a width of \SI{1}{\micro\meter} and a diffusion length of \SI{0.15}{\micro\meter}. (b) The distribution of carrier recombination positions when illuminating the bottom of the same NW. (c) The variation of the carrier occupation factors at the surface, in the bulk volume and at the bottom interface when illuminating the top of NWs with different widths. The mean separation between the carrier recombination positions and the base is also shown. This calculation is for NWs with a diffusion length of \SI{0.15}{\micro\meter} and a surface thickness of \SI{10}{\nano\meter}. (d) The carrier occupation factors and mean separation when illuminating the bottom of the same NWs of different widths. (e) The impact of changing the diffusion length when illuminating the top of a NW with a width of \SI{1}{\micro\meter} and a surface thickness of \SI{10}{\nano\meter}. (f) The impact of changing the diffusion length when illuminating the top of a NW with a width of \SI{1}{\micro\meter} and a diffusion length of \SI{0.15}{\micro\meter}.}
    \label{fig:diffusion}
\end{figure*}

The measured properties of the NWs depends on where in the NW the photogenerated carriers recombine. This recombination location affects the recombination energy, and thus the measured bandgap, and impacts the recombination mechanism for the carriers (i.e. non-radiative at traps or radiative). The distribution of recombination locations is determined by the NW geometry and size, the carrier diffusion length and the effective thickness of the surface layers. This is a complex 2-dimensional diffusion problem, with no trivial analytical solution.

The recombination distribution is calculated for a NW using a Monte-Carlo approach. In this method, 40,000 carriers are injected individually at random points on the top surface of a NW cross section, which is equivalent to exciting carriers at the NW surface. A 2D grid of positions with \SI{5}{\nano\meter} spacing are generated to represent sites which carriers can occupy. After each incremental time step, two random numbers are generated to determine the behaviour of each carrier. The first number is checked against the carrier recombination probability to establish if the carrier recombines. If recombination occurs, the carrier is removed and the recombination position is recorded. The second number determines the nearest neighbour site where the carrier moves. This process continues until each carrier recombines. This produces a distribution of positions in the cross section where the carriers have recombined, an example is shown in Figure~S2a. The process is repeated for illumination from the bottom by randomly placing carriers along the bottom interface, providing a carrier distribution as shown in Figure~S2b.

The Monte-Carlo simulations are repeated for a range of NW widths, carrier diffusion lengths and surface thicknesses. In each case, the proportion of carriers that are within the surface thickness from the top surface and bottom interface are calculated. The remaining carriers are in the bulk volume of the NW. These occupation factors are used, along with the geometrical volumes, to calculate the carrier densities in each of these regions. Additionally, the mean separation between the carriers and the base of the NW is calculated, which influences the effective recombination bandgap due to strain effects.

Figure~S2c,d shows that, for a fixed surface thickness and diffusion length, when illuminating from the top the majority of carriers recombine in the NW bulk volume (\(\approx\) \SI{70}{\percent}), with a reduced amount at the top surface (\(\approx\) \SI{20}{\percent}) and even fewer from the bottom interface (\(\approx\) \SI{10}{\percent}). This provides further justification to the assertion that the PL time decays show recombination mainly from the bulk volume, and that top surface recombination is only observed when exciting from the top of the NW. For narrow NWs the proportion of carriers in the bulk reduces due to a reducing NW volume. Reducing the NW width also reduces the carrier-base separation in a linear fashion.

When illuminating from the bottom, the majority of carriers again recombine in the bulk volume (\(\approx\) \SI{70}{\percent}), but more carriers recombine at the bottom interface (\(\approx\) \SI{20}{\percent}) than the top surface (\(\approx\) \SI{10}{\percent}). In these experiments there is a large contribution from carrier recombination in the bulk volume, and a reduced contribution from recombination at the top surface.

Figure~S2e shows that, for diffusion lengths longer than the NW height (\SI{0.5}{\micro\meter} in this case), the carrier distribution in the NW is uniform, and so the output parameters of the model are unaffected by increasing the diffusion length. If the diffusion length is less than the NW height, carriers can no longer diffuse throughout the NW and the distribution is non-uniform, as illustrated in Figure~S2a. This causes a higher proportion of carriers to be trapped at the surface, close to where they are generated, reducing the number of carriers in the bulk volume and at the bottom interface, and increasing the average carrier-base separation. 

Figure~S2f shows that changing the effective surface thickness causes a trivial change in the occupation factors: a thicker surface increases the proportion of carriers at the top surface and bottom interface, and reduces the number of carriers in the bulk.

The carrier diffusion length and surface thickness are fitting parameters in the carrier recombination model, which aims to fit data from $>$8,000 NWs with different widths. Due to limits in computational power, it is not feasible to include the full Monte-Carlo diffusion calculation in the recombination model. To circumvent this, a machine learning (ML) model is trained on the results from 13860 simulations of NWs with different widths, diffusion lengths and surface thicknesses. This ML model is capable of predicting the occupation factors and carrier-base separations with an error of $<$\SI{3}{\percent}. Crucially, the ML model takes \(\approx\) \SI{90}{\micro\second} to make these predictions, and can therefore be incorporated into a nonlinear fitting routine.

\subsection*{Photoluminescence Model}

\begin{figure*}
    \centering
    \includegraphics[width=1\linewidth]{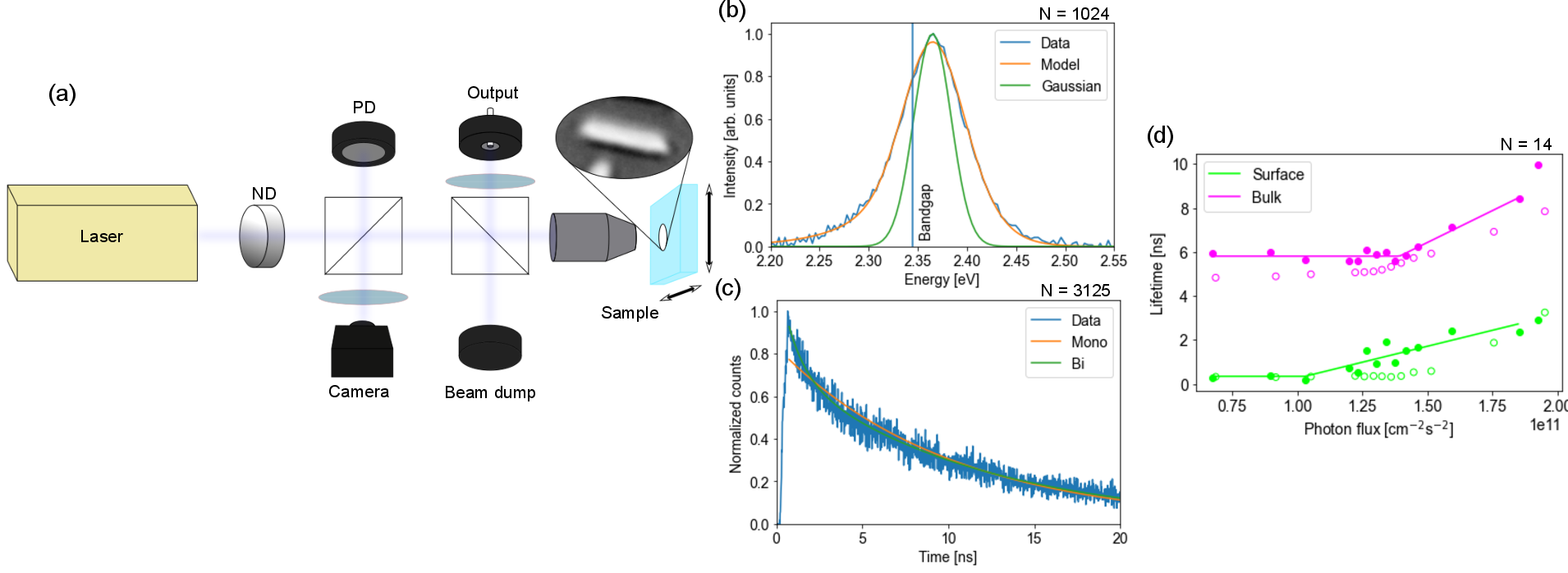}
    \caption{Optical spectroscopy and modelling of individual nanowires. (a) A schematic of the home-made microscope. Laser excitation passes through a neutral density filter (ND) and light reflects off a 50:50 beamsplitter onto a photodiode (PD) used to monitor the beam power. Transmitted light is sent to an objective lens and sample stage. Light from the sample is collected by the objective lens, is reflected by a beamsplitter, before being focused into an optical fibre. Light from the sample is also focused onto a machine vision camera for imaging. (b) Room temperature PL spectrum fit by Equation~S2. The optical bandgap is indicated, which lies between the minimum and peak emission energies due to inhomogeneity in the NW. The Gaussian broadening term is also shown, which is the largest contributor to the FWHM of the emission. (c) A room temperature PL time decay measurement, showing a mono-exponential and bi-exponential fit to the data using Equation~S4. The squared sum of the residuals are 13.7 and 12.8 for mono and bi-exponential decays respectively. in this case the bi-exponential fit is the most appropriate. (d) The power dependence of the PL lifetimes extracted from PL time decay fits. The full circles show the results from a typical NW and the open circles are taken from the NW with the largest fractional change in lifetime. The full circles have been fit with Equation~S5 for the surface and bulk decays.}
    \label{fig:PL_TCSPC}
\end{figure*}

The room temperature photoluminescence spectra were measured using a home-built optical microscope, a schematic is shown in Figure~S3a. These spectra were fit with a model which assumes that photogenerated carriers rapidly cool to the conduction and valence band edges and occupy a thermal distribution. This is reasonable for CsPbBr\textsubscript{3}, as the cooling times are significantly faster than the recombination times~\cite{Boehme2020}. Exciton effects can also be safely ignored at these temperatures due to a small exciton binding energy~\cite{Baranowski2020}. If it is assumed that the thermal energy is large enough at room temperature such that the Fermi-Dirac carrier distribution can be approximated by a Maxwell-Boltzmann distribution, the PL spectrum can be estimated by Equation~S2:
\begin{equation}
\label{equ:PL}
    I_{PL}(E) = G(E,\sigma) \otimes B(E) \text{exp}\left(-\frac{E-E_{\rm{0}}}{k_\beta T}\right)
\end{equation}
where \(E_{\rm{0}}\) is the average bandgap energy and \(B(E)\) is the joint density of states of the carriers. 

In Equation~S2, \(G\) is a Gaussian function with standard deviation \(\sigma\). This approximates the inhomogeneity in the NW via a convolution. A large source of this inhomogeneity is the strain, which varies with distance from the surface~\cite{Oksenberg2018Surface-GuidedResponsea}. The strain modifies the bandgap and so carriers which recombine closer to the NW/substrate interface will have a smaller PL peak energy.

As these NWs have widths much greater than the Bohr radius~\cite{Berestennikov2019}, the density of states is 3-dimensional. A modification to this is required at emission energies below the bandgap due to an exponential tail on the PL spectra. This is due to disorder on the atomic level in the NWs giving rise to a small density of states below the bandgap known as an Urbach tail, and is defined by an Urbach energy, \(E_U\)~\cite{Ledinsky2019}. As a result, the density of states is given by Equation~S3:
\begin{equation}
\label{equ:DoS}
  B(E)=\begin{cases}
    A_{\rm{1}}\sqrt{E - (E_{\rm{0}}-dE)}, & \text{if $E>E_{\rm{0}}-dE$}.\\
    A_{\rm{2}}\text{exp}\left(\frac{E-E_{\rm{0}}}{E_{\rm{U}}}\right), & \text{if $E<E_{\rm{0}}-dE$}.
  \end{cases}
\end{equation}
where \(A_{\rm{1}}\) and \(A_{\rm{2}}\) are the amplitude of each component and \(dE\) is a small offset energy required to connect the two energy regimes.

An example of a fit to a room temperature PL spectrum from a NW is shown in Figure~S3b.  The average bandgap is \(\approx\)\SI{15}{meV} below the peak emission energy, and is \(\approx\)\SI{100}{meV} above the minimum emission energy. This is due to the large degree of inhomogeneity in these samples. This inhomogeneity is the major cause of the width of the PL peak, as illustrated by the Gaussian.

The Urbach tail is illustrated at the lowest energies in Figure~S3b. The Urbach energy is independent of wire width: it has a Gaussian distribution centred on \SI{12(2)}{\milli\electronvolt}, and is comparable to the Urbach energy of CsPbBr\textsubscript{3} single crystals, where an upper limit of \SI{19}{\milli\electronvolt} was previously determined by absorption measurements~\cite{Rakita2016}.

\subsection*{TCSPC Model}

The room temperature PL time decays for each NW were measured using the same microscope system. These decays were either mono-exponential or bi-exponential in shape. All decays were fit with both a mono-exponential and a bi-exponential fit, given by Equation~S4:
\begin{equation}
\label{equ:TCSPC}
  I(t)=\begin{cases}
    A_{\rm{1}}\text{exp}\left(\frac{-t}{\tau_{\rm{1}}}\right)\\
    A_{\rm{1}}\text{exp}\left(\frac{-t}{\tau_{\rm{1}}}\right) + A_{\rm{2}}\text{exp}\left(\frac{t}{\tau_{\rm{2}}}\right)
  \end{cases}
\end{equation}
where \(A\) is an amplitude factor and \(\tau\) is the recombination lifetime for this mechanism. An example of this fitting is shown in Figure~S3c. To determine the overall shape of the decay, the reduced chi-squared for each fit was compared, and the fit with the smallest chi squared was selected. The integral under the fit was also calculated - which can be used to infer the number of carriers recombining at the surface and in the bulk.

The laser excitation power was attenuated using an ND filter to measure the power dependence of the lifetimes. This experiment was performed on 220 NWs when exciting from the top. The same fitting procedure was applied to PL time decays at different excitation powers, to determine the carrier density dependence of the PL lifetimes. As shown in Figure~S3d, the recombination lifetimes increase above a threshold, \(P_{\rm{Thresh}}\), which differs at the surface and the bulk. To account for this, a linear increase in the lifetimes was fit to the data using Equation~S5:
\begin{equation}
\label{equ:TCSPC_pow}
  \tau(P)=\begin{cases}
    C, & \text{if $P<P_{{\rm{Thresh}}}$}.\\
    C + m(P- P_{{\rm{Thresh}}}), & \text{if $P>P_{{\rm{Thresh}}}$}
  \end{cases}
\end{equation}
where \(m\) is the slope of the variation, and \(C\) is a lifetime offset. This linear increase is suitable because, as demonstrated in Figure~S3d, we do not reach the photon flux required to saturate all of the carrier traps.

\subsection*{High throughput results}
\label{sec:high}

\begin{figure*}
    \centering
    \includegraphics[width=1\linewidth]{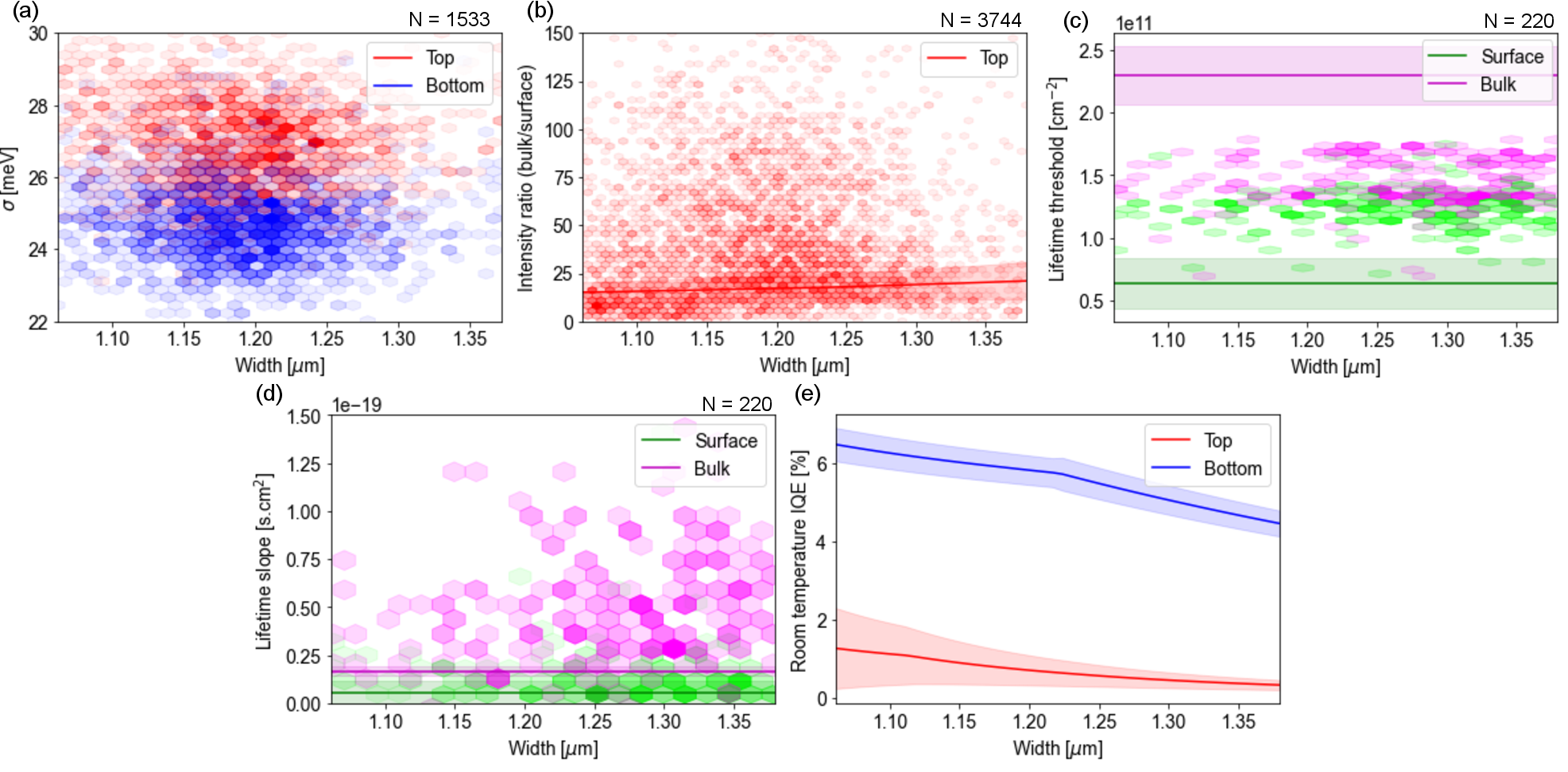}
    \caption{ High throughput spectroscopy results of the NW population. (a) The variation of the PL inhomogeneity parameter \(\sigma\) for NWs of different widths, when excited from the top and bottom. (b) A histogram of the ratio of the integral of the TCSPC decays as determined by Equation~S6 for top illumination. (c) The flux threshold above which the PL lifetime increases for both bulk and surface recombination, taken from fits to data from 220 NWs such as those in Figure~S3d. (d) The slope of the lifetime variation with increasing photon flux above threshold, as described by Equation~S10. (e) The variation of the recombination IQE for NWs of different widths, when excited from the top and bottom, as predicted by the model. }
    \label{fig:recomb}
\end{figure*}

This section covers other results from the recombination model linking the PL, TCSPC and power dependent measurements. All fits shown are generated from the same data-constrained model as presented in the main paper and the quoted uncertainties are SEs.

The fits to the PL spectra provide the parameter \(\sigma\), taken from the Gaussian convolution, which is a measure of the degree of inhomogeneity in the wire cross section. \(\sigma\) was determined for 1533 NWs when illuminating from the top and bottom. The variation of \(\sigma\) with NW width is shown in Figure~S4a. These results show that \(\sigma\) is consistently larger for top illumination, with mean values of \SI{27}{\milli\electronvolt} and \SI{25}{\milli\electronvolt} for top and bottom illumination. This increased homogeneity when from the top is further evidence for the bandgap gradient illustrated in Figure~1d,e: due to the triangular cross section, carriers generated near the air/NW interface are more likely to recombine in areas of different bandgap, increasing the value of \(\sigma\) when compared to carriers generated near the NW/substrate interface.

There is also a slight negative correlation between \(\sigma\) and NW width. The Pearson's correlation coefficients (\(r\),\(p\)) are -0.1, 5E-5 and -0.08, 0.002  for top and bottom illumination respectively. Therefore, the degree of inhomogeneity increases for thinner NWs. This reflects a larger bandgap gradient across the NW cross section, which is a consequence of increased lattice rotation effects in the thinner NWs, as determined from the PL measurements in Figure~2a.

The fits to the TCSPC measurements can also be used to constrain the carrier populations in the surface and the bulk. The integral of the separate mono-exponential components in the PL time decays, \(I_{\rm{V}}\) and \(I_{\rm{S}}\), are related to the carrier populations by Equation~S6:
\begin{equation}
    \label{equ:Int_ratio}
    \frac{I_{\rm{V}}}{I_{\rm{S}}} = \frac{N_{\rm{V}}}{N_{\rm{S}}} =\frac{B_{\rm{V}}}{B_{\rm{S}}}
\end{equation}
where \(B_{\rm{V,S}}\) are the occupation factors for the bulk volume and surface. These values depend on the NW width, carrier diffusion length and effective surface thickness, and are calculated using the Monte-Carlo diffusion model. This fit is shown in Figure~S4b. There is no trend with optical width, due to the large degree of scatter in the data.

The power dependence of the PL lifetimes can also be used to constrain the model. The measurements shown in Figure~S3d were repeated for 220 NWs: Figure~S4c,d shows that both the threshold power density and lifetime slope are higher for bulk recombination, when compared with the surface, and that neither of these vary with NW width. 

These results can be explained by considering the density of unoccupied traps (\(n_{{\rm{t,eff}}}\)) and free carriers, which are related to the trap density and the photon flux. At high photon fluxes, the trap states begin to fill up, reducing \(n_{{\rm{t,eff}}}\): this is shown as an increase in the lifetime in Figure~S3d. Both the threshold flux and the rate of change of the lifetime will be dependent upon the trap density at the surface and in the bulk. Equation~S7,8 assume that the threshold carrier density is proportional to the trap density:
\begin{align}
\label{equ:threshold}
n_{{\rm{V,threshold}}} = \alpha_{\rm{V}} n_{{\rm{tV}}} \\
n_{{\rm{S,threshold}}} = \alpha_{\rm{S}} n_{{\rm{tS}}}
\end{align}
where \(\alpha\) is a proportionality factor. The threshold can be related to the threshold flux using Equation~2 in the main paper. The effective density of traps is then reduced by the density of carriers above this threshold. Equation~S9 describes this for the bulk volume and top surface: 
\begin{equation} 
\label{equ:effectiveV}
  n_{{\rm{t,eff}}}=\begin{cases}
    n_{{\rm{t}}}, & \text{if $n < \alpha n_{{\rm{t}}}$}.\\
    (1 + \alpha) n_{{\rm{t}}} - n, & \text{if $n > \alpha n_{{\rm{t}}}$}.
  \end{cases}
\end{equation}

The rate of change of the lifetime can be calculated by differentiating Equation~5,6 with respect to the photon flux, resulting in Equation~S10:
\begin{equation}
\label{equ:slope}
\frac{dt}{dn_{{\rm{flux}}}} = \frac{\alpha k_{{\rm{t}}} n_{{\rm{t}}} \langle abs \rangle\ B \langle w_{{\rm{spot}}} \rangle\ t^2 }{V}
\end{equation}

The fits to the lifetime thresholds (Figure~S4c) and lifetime slopes (Figure~S4d) are the weakest part of the model. This is because these datasets have an order of magnitude fewer objects than the others, and therefore have a reduced weight in the fitting routine. The surface slope fit is within 1~$\sigma$ of the data. The bulk slope fit fails to account for the large spread in the data, resulting in a fit that is 1.1~$\sigma$ away from the data. The threshold fits provide the least accurate values that lie between 2.2 and 3.3~$\sigma$ from the data. Nevertheless, these fits are a useful way to further constrain the global model.

The model can calculate the recombination IQE for different widths: this is shown in Figure~S4e. When exciting from the top, the mean IQE extracted from the model is \SI{0.7(1)}{\percent}. The IQE reduces with increasing NW width, due to a reduction in the radiative rate and an increase in the surface non-radiative rate. As shown in Figure 3, this is due to a reduction in carrier density and a reduction in the number of saturated traps at the surface. When the NW is excited from the bottom, the mean IQE increases to \SI{5.8(4)}{\percent} due to reduced top surface recombination.

\begin{table}
\centering
\label{table:model}
\begin{tabular}{lllc}
\hline
Dataset                  & NW region & Illumination orientation & \multicolumn{1}{l}{Equation} \\ \hline
PL bandgap               &           & Top                      & 7                           \\
PL bandgap               &           & Bottom                   & 7                           \\
TCSPC lifetimes          & Bulk      & Top                      & 5                           \\
TCSPC lifetimes          & Surface   & Top                      & 6                           \\
TCSPC lifetimes          & Bulk      & Bottom                   & 6                           \\
TCSPC lifetime threshold & Bulk      & Top                      & S7                           \\
TCSPC lifetime threshold & Surface   & Top                      & S8                           \\
TCSPC lifetime slope     & Bulk      & Top                      & S10                           \\
TCSPC lifetime slope     & Surface   & Top                      & S10                          \\
IQE     &    &                       & 8    
   \\
Mobility    &    &                       & 9   
\\ \hline
\end{tabular}
\caption{A summary of the multi-modal data-sets produced by the high-throughput spectroscopy methodology, and the equations which describe the component of the recombination model which fits to each data-set.}
\end{table}
As shown in Figure~2 and S4, the model can fit the full multi-modal data-set. A summary of the data-sets involved is provided in Table~S1, and the output parameters are summarised in Table~S2. A flowchart showing the major connections in the model is shown in Figure~S5.

\begin{figure*}
    \centering
    \includegraphics[width=1\linewidth]{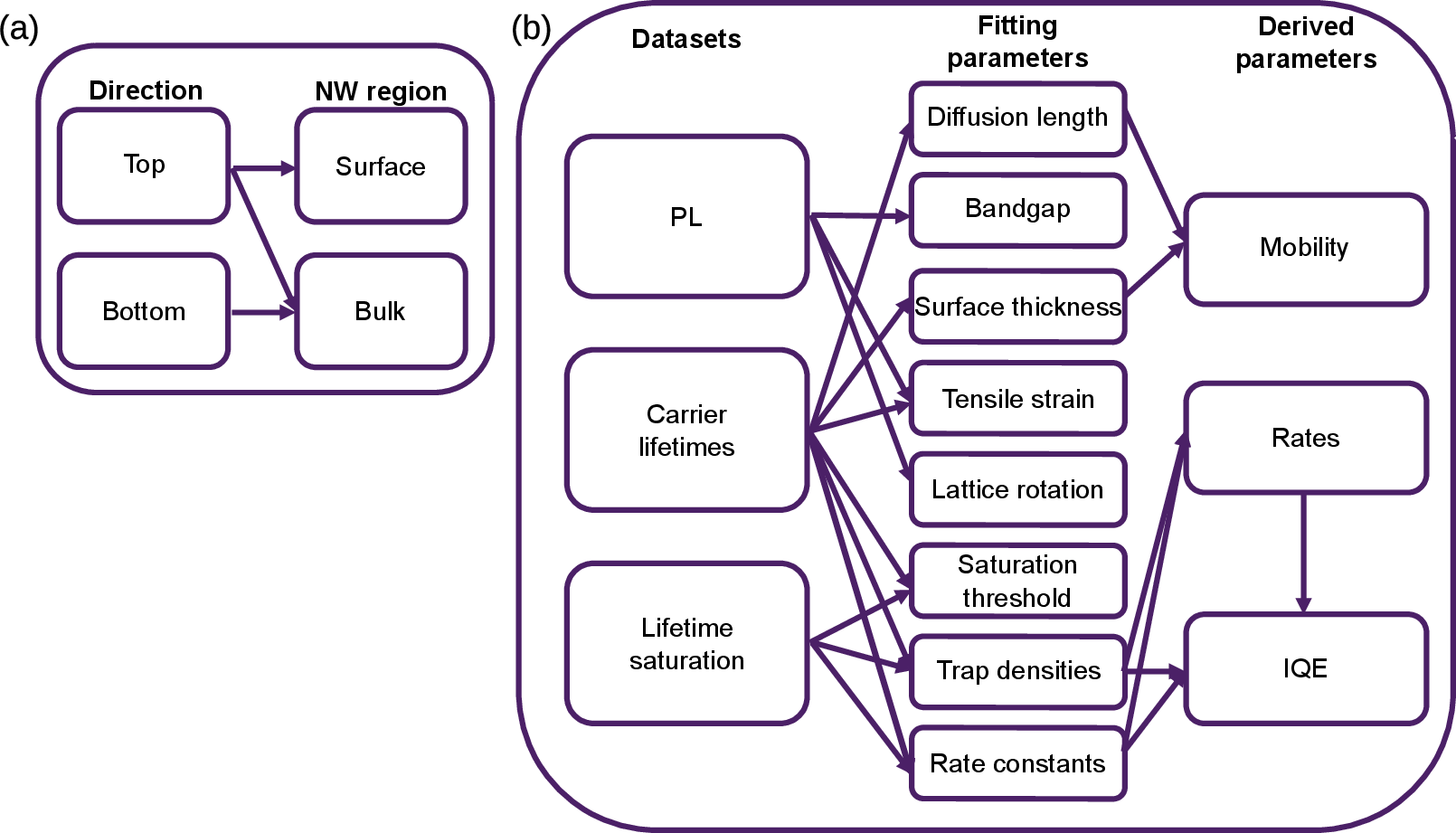}
    \caption{A flowchart summary of the holistic model. (a) The sensitivity of the measurement to different NW regions, depending upon the illumination direction. (b) The major connections between datasets, fitting parameters and derived parameters.}
    \label{fig:flow}
\end{figure*}

The diffusion length of \SI{0.25(2)}{\micro\meter} is smaller than the best in class values obtained in previous studies~\cite{Stranks2013a,Yettapu2016a,Lai2018,Oksenberg2021a}. This may reflect the holistic nature of this study, which involves the entire NW population, rather than focusing on the best performing. The effective surface thickness is \SI{6(1)}{\nano\meter}, which is roughly 10 atomic layers and is comparable to the exciton Bohr Radius in this material (\SI{7}{\nano\meter}~\cite{Protesescu2015}).

The surface trap density is \SI{7.1(3)E16}{\per\centi\meter\cubed}, which is comparable to trap density in the bulk (\SI{8.6(4)E16}{\per\centi\meter\cubed}). Despite this, the calculated non-radiative rate at the surface is higher than in the bulk. This is partially due to the fact that the surface traps lie within \SI{6(1)}{\nano\meter} of the surface, leading to a high areal density of \SI{4.3(7)E10}{\per\centi\meter\squared}. The surface rate constants are also higher: The non-radiative rate is \SI{1.20(5)E-24}{cm^6s^{-1}}, which is more than an order of magnitude larger than the bulk non-radiative rate of \SI{4.4(2)E-26}{cm^6s^{-1}}. This difference may mean that the surface traps are more likely to capture carriers. This is likely because the carriers are initially generated at the surface layer and are in close proximity to the surface traps. The rate for bulk recombination will be reduced as carriers must first diffuse to these states, and hence trapping is less likely.

The radiative recombination rate is low in comparison to the nonradiative rates, this is compatible with previous studies~\cite{Jiang2019a}, and is expected for these NWs where carrier traps are important.

The threshold trap density factor is a measure of the relative proportion of carriers, relative to the trap density, that are required to start to saturate the traps. The value for the bulk is \SI{0.36(2)}, and it is \SI{0.23(1)} at the surface. This suggests that a higher proportion of carriers in the bulk are needed to saturate the traps: this result may be linked to the recombination rate of trapped carriers, suggesting that this is higher in the bulk.

\begin{table}
\centering
\label{table:fitting parameters}
\begin{tabular}{llcc}
\hline
Parameter       & Description                                            & Fit value & unit       \\ \hline
\(L_{{\rm{D}}}\)                  & Diffusion length of minority carriers     & \SI{0.25(2)}{} &  {[}$\mu$m{]}   \\
\(L_{{\rm{surface}}}\)               & Effective thickness of surface layer          & \SI{6(1)}{} & {[}nm{]}    \\
\(k_{\rm{rV}}\)                     & Radiative recombination rate in the bulk   & \SI{7.5(3)E-27}{} & {[}cm$^6$s$^{-1}${]}    \\
\(k_{{\rm{tV}}}\)                     & Nonradiative rate in the bulk                 & \SI{4.4(2)E-26}{}  & {[}cm$^6$s$^{-1}${]}   \\
\(k_{{\rm{tS}}}\)                     & Nonradiative rate at the surface               & \SI{1.20(5)E-24}{}  & {[}cm$^6$s$^{-1}${]}   \\
\(N_{{\rm{tV}}}\)                     & Trap density in the bulk                      & \SI{8.6(4)E16}{} & {[}cm$^{-3}${]}   \\
\(N_{{\rm{tS}}}\)                     & Trap density at the surface                    & \SI{7.1(3)E16}{}  & {[}cm$^{-3}${]} \\
\(\alpha_{{\rm{V}}}\) & Threshold trap density factor in the bulk                & \SI{0.36(2)}{} &  \\
\(\alpha_{{\rm{S}}}\) & Threshold trap density factor at the surface             & \SI{0.23(1)}{} &    \\
\(E_{\rm{g}}\)                   & Unstrained bandgap of the NW                    & \SI{2.37(10)}{} & {[}eV{]}   \\
\(T\)    & Bandgap shift due to tensile strain  & \SI{64(3)}{} & {[}meV{]} \\
\(L_{\rm{rot}}\)   & Variation of bandgap due to lattice rotation  & \SI{0.26(1)}{} & {[}$\mu$m{]}\\
\(A\)    & Amplitude of bandgap due to lattice rotation  & \SI{1.0(1)}{} & {[}meV{]} \\ \hline
\end{tabular}
\caption{A summary of the fitting parameters used in the recombination model which was applied to the multi-modal data-set.}
\end{table}

The unstrained bandgap was extracted to be \SI{2.4(1)}{\electronvolt}, consistent with the literature~\cite{Mannino2021a}, albeit with a large uncertainty. The tensile strain at the base was determined to be \SI{64(3)}{\milli\electronvolt}. Literature values suggest that the bandgap shift with strain is \SI{22}{\milli\electronvolt\per\percent}~\cite{Li2019}, therefore this model estimates that the interfacial strain is \SI{2.9(2)}{\percent}. This is large in comparison with the strain calculated from bulk lattice parameters (\SI{0.7}{\percent}~\cite{Oksenberg2020a}). However, the thermal expansion coefficient of the CsPbBr\textsubscript{3} is an order of magnitude larger than the sapphire, which will likely lead to additional strain during post-growth cooling~\cite{Oksenberg2020a}. The NW relaxes exponentially with distance from the interface with a characteristic length of $L_{\rm{rot}}=$\SI{0.26(1)}{\micro\meter}, and amplitude of \SI{1.0(1)}{\milli\electronvolt}.

\subsection*{Temperature dependence of the photoluminescence}
\label{sec:Tdep}

\begin{figure*}
    \centering
    \includegraphics[width=1\linewidth]{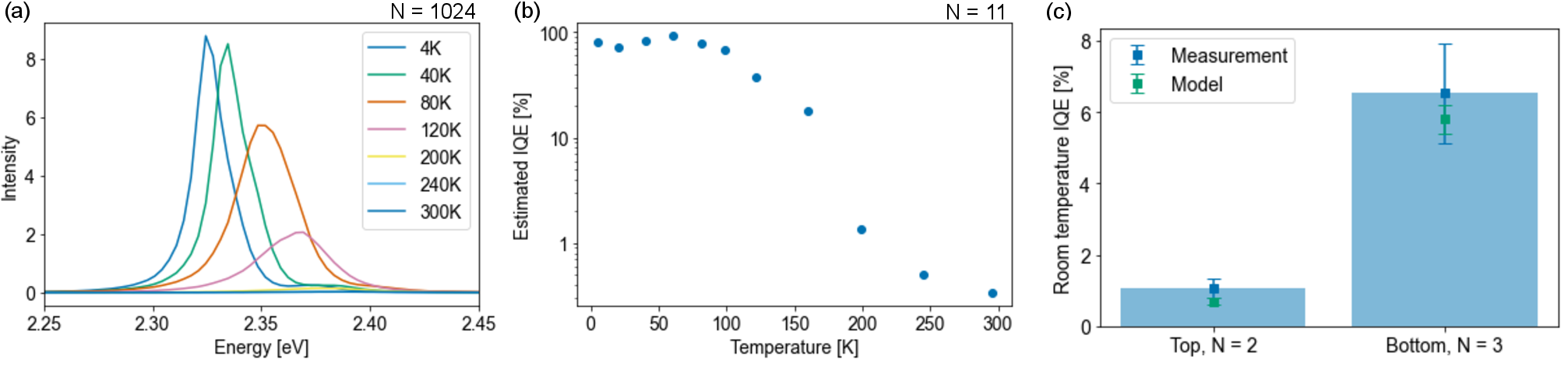}
    \caption{Temperature dependent PL results of a small number of NWs. (a) PL spectra at different temperatures for a single NW. (b) The estimated IQE of the same NW at different temperatures. (c) The weighted mean of the room temperature estimated IQE of NWs when exciting from the top and bottom.}
    \label{fig:Tdep}
\end{figure*}

Temperature dependent PL was performed for NWs by mounting the substrate on the cold finger of a Montana cryostation. The same laser and spectrometer, as used in the main text, were also used to investigate the samples. The sample temperature was varied between \SI{10}{\kelvin} and \SI{300}{\kelvin}, refocusing the excitation spot at each temperature. The measurements were performed when exciting the NWs from the top and bottom, looking at 2 NWs from the top and 3 from the bottom. 

The PL spectra at different temperatures for a single NW when exciting from the top are shown in Figure~S6a. At low temperature, the majority of carriers will form excitons, and recombination will be fast and highly radiative~\cite{Wolf2018a}. As the temperature is increased, the emission blueshifts by \SI{42}{\milli\electronvolt} due to photon-phonon coupling~\cite{Wolf2018a}. The excitons will also dissociate and the recombination efficiency will decrease~\cite{Yettapu2016a}, and shallow trap states will be activated~\cite{Baranowski2020}. We assume that the absorption strength and hence excitation density does not vary with temperature. This is likely valid for the excitation conditions used here (\SI{405}{\nano\meter}) which is significantly above the band-edge and is therefore not strongly affected by excitonic absorption~\cite{Diroll2018}. The ratio of the integrated intensity of the PL spectrum at low temperatures and elevated temperatures can therefore be used to estimate the IQE. This calculation was performed for the five NWs: the results for a single NW at different temperatures are shown in Figure~S6b.

The room temperature estimated IQE was calculated for each NW, and a weighted mean was calculated: this is summarised in Figure~S6c. When exciting from the top, the mean IQE was \SI{1.1(3)}{\percent} - this is comparable to the predictions from the model, shown in Figure~S4e. Furthermore, when exciting from the bottom of the NWs, the mean IQE increased to \SI{7(1)}{\percent}.These results are both consistent with and therefore verify the capability of the model to predict optoelectronic properties of the NWs.

\medskip

\providecommand{\latin}[1]{#1}
\makeatletter
\providecommand{\doi}
  {\begingroup\let\do\@makeother\dospecials
  \catcode`\{=1 \catcode`\}=2 \doi@aux}
\providecommand{\doi@aux}[1]{\endgroup\texttt{#1}}
\makeatother
\providecommand*\mcitethebibliography{\thebibliography}
\csname @ifundefined\endcsname{endmcitethebibliography}
  {\let\endmcitethebibliography\endthebibliography}{}

\end{document}